\documentclass[aip,twocolumn,jcp]{revtex4}
\usepackage{amsmath,amssymb,graphicx}
\usepackage{epsfig}
\usepackage{graphicx}
\usepackage{epstopdf}
\usepackage{enumerate}
\usepackage{comment}

\let\baraccent=\= % rename builtin command \= to \baraccent

\newcommand{\beq}{\begin{equation}}
\newcommand{\eeq}{\end{equation}}

\newcommand{\bea}{\begin{eqnarray}}
\newcommand{\eea}{\end{eqnarray}}

\begin{document}

\title{
Electron transport in nanoscale junctions with local anharmonic modes}
\author{Lena Simine and Dvira Segal}
\affiliation{Chemical Physics Theory Group, Department of Chemistry, University of Toronto,
80 Saint George St. Toronto, Ontario, Canada M5S 3H6}

\date{\today}
\begin{abstract}
We study electron transport in nanojunctions in which an electron on a quantum dot or a molecule is interacting
with an $N$-state local impurity, a harmonic (``Holstein") mode, or a two-state system (``spin").
These two models, the Anderson-Holstein model and the spin-fermion model,
can be conveniently transformed by a shift transformation into
a form suitable for a perturbative expansion in the tunneling matrix element.
We explore the current-voltage characteristics of the two models in the limit of high
temperature and weak electron-metal coupling using a kinetic rate equation formalism,
considering both the case of an equilibrated impurity, and the unequilibrated case.
Specifically, we show that the analog of the Franck-Condon blockade physics is missing in the
spin-fermion model.
We complement this study by considering the low-temperature quantum adiabatic limit of
the dissipative spin-fermion model, with fast tunneling electrons and a slow impurity.
While a mean-field analysis of the Anderson-Holstein model suggests that nonlinear functionalities,
bistability and hysteresis may develop, such effects are missing
in the spin-fermion model at the mean-field level.
\end{abstract}

%\pacs{05.70.-a, 05.70.Ln, 89.70.-a, 89.70.Kn}

\maketitle

%-------------------

\section{Introduction}

% general intro
Molecular electronic devices have been of significant interest in the past decade
offering a fertile playground for studying fundamentals of nonequilibrium many-body
physics \cite{NatelsonRev,Latha,40years}. The simplest junction includes
a single molecule, possibly gated, bridging two voltage-biased leads.
Mechanisms of charge transport in such systems, specifically,
the role of many-body interactions (electron-phonon, electron-electron, electron-magnetic impurity)
can be resolved e.g., from direct current-voltage measurements,
studies of current noise, and from different types of spectroscopy,
 inelastic electron tunneling spectroscopy and Raman studies
% to obtain the nonequilibrium vibrational distribution in the junction
\cite{NatelsonRev,Latha,40years}.
%
% vibrations
Naturally, molecular electronic degrees of freedom are
coupled to nuclear vibrations, and signatures of this interaction
appear through peaks in the differential conductance \cite{IETS},
nonequilibrium heating of vibrational modes
\cite{NatelsonVib},
the presence of the Franck-Condon blockade \cite{KochFCB,KochFCBlong,Kochvib,FCBexp} % check koch papers
and other (proposed) effects:
vibrational instabilities \cite{Mitra, KochNitzan,ThossInst},
vibrationally induced negative differential resistance \cite{ThossVib},
current hysteresis, switching
and bistability  \cite{NitzanBi,NitzanBiGF,Brat1,Brat2, Kosov,RabaniBi1,RabaniBi2},
and electron-pair tunneling \cite{KochPair}.

% AH
In the simplest theoretical description of electron-conducting junctions
only degrees of freedom that immediately participate
in the transport process are included. The single-impurity ``Anderson-Holstein" (AH) model
comprises a single electronic level (dot) and a local harmonic-vibrational mode.
Electrons on the dot may electrostatically repel, but the metals are treated as Fermi gases
with noninteracting electrons.
This minimal model has been revisited many times, and it has been examined in different limits by means of analytical, perturbative and numerical techniques.
Perturbation expansions were performed in either the electron-phonon interaction parameter or the
tunneling matrix element to the metals, resulting in
Redfield \cite{Redfield,ThossInst,ThossVib}, polaronic \cite{Mitra, KochFCBlong,Kochvib,KochThermoE} and
Keldysh Green's function equations of motion \cite{GF1}.
Numerically exact tools provide
transient effects towards the steady-state limit. Among such techniques we list wave-function based methodologies \cite{MCTDH, RabaniBi1,RabaniBi2},
time-dependent numerical renormalization group approaches \cite{RGAnders,Schiller}, and
iterative-deterministic  \cite{ISPIAH} and diagrammatic Monte Carlo
\cite{MillisMC2,Lothar} path-integral tools. %, RabaniMC

%%-----------------------------
% schemes
\begin{figure}[htbp]
\hspace{4mm}
\vspace{-1mm} {\hbox{\epsfxsize=150mm \hspace{0mm}\epsffile{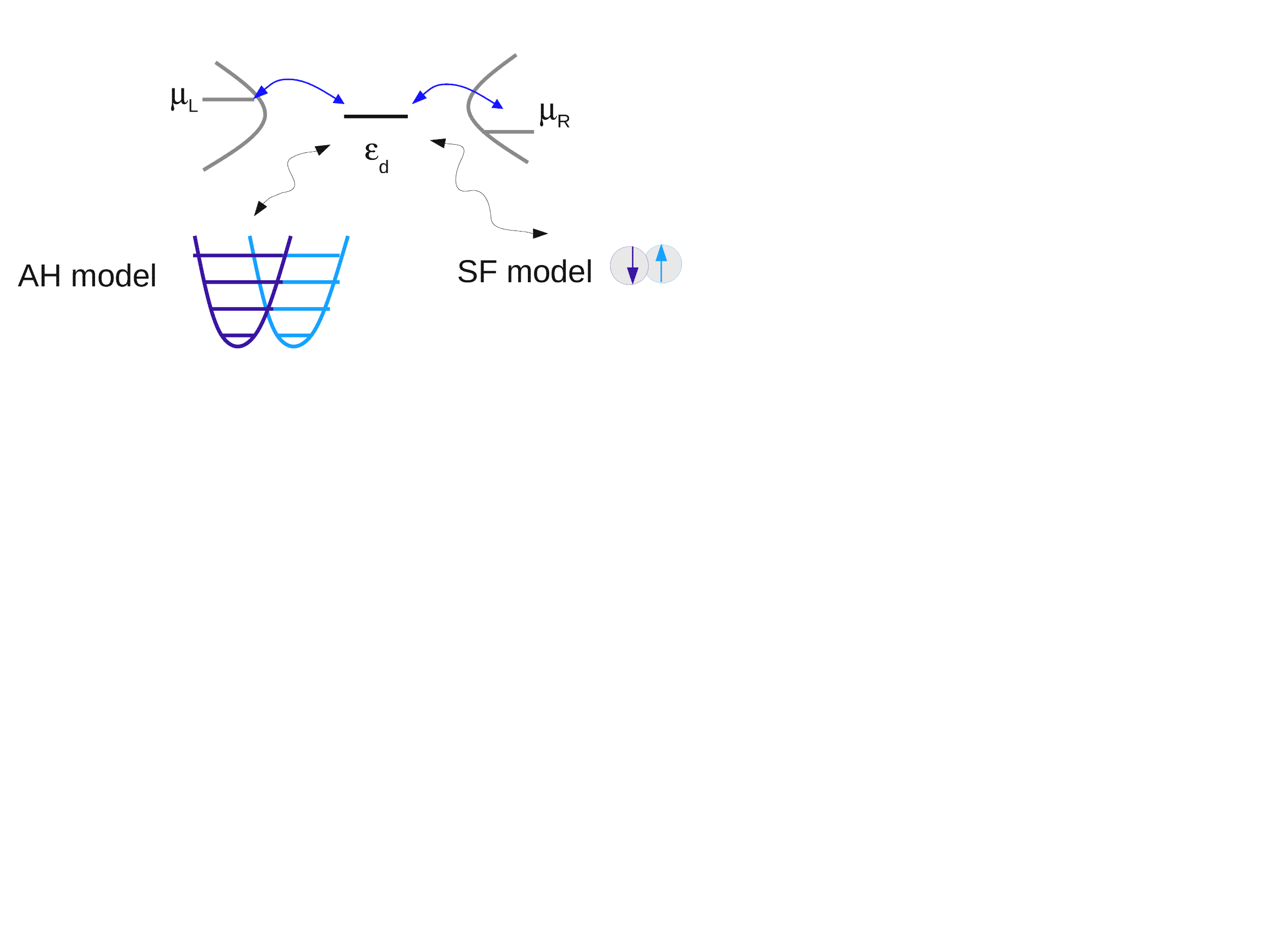}}}
\caption{\normalsize
{Minimal modeling of nanojunctions with a single electronic level
 (energy $\epsilon_d$) coupled to two metals.
In the Anderson-Holstein (AH) model the vibrational mode is displaced
depending on the charge number in the dot.
The spin-fermion model (SF) is a truncated version of the AH model.
Its (nondegenerate) two states describe e.g.,
 an anharmonic mode or a magnetic impurity
in an external magnetic field. Electrons residing on the dot may flip
the spin state.
}}
\label{Scheme}
\end{figure}
%-----------------------------------

% beyond the harmonic limit
The Anderson-Holstein model describes the potential energy of atoms displaced from equilibrium
within the harmonic approximation.
It is important to examine nanojunctions beyond this ideal limit, and describe more realistic structures.
Several recent studies considered the role
of molecular anharmonicity (in the form of a Morse potential) on charge transport characteristics,
generally displaying small effects \cite{KochNitzan,KochMorse}. % XXX
More fundamentally, the AH model should be extended beyond the harmonic limit to describe
 situations in which electrons on the dot couple to naturally anharmonic
degrees of freedom, intramolecular, or in the surrounding.
Such situations  arise in different setups:
nanojunctions consisting local magnetic impurities \cite{Park,SpinRev},
 nanoelectromechanical devices \cite{Ralf},
semiconductor quantum dots coupled to nuclear spins in the surroundings \cite{Brandes0,Brandes1,Brandes2},
charge sensing in the junction through e.g., nitrogen-vacancy centers \cite{NV1,NV2},
and when the electronic degrees of freedom are coupled to
(discrete or continuous) molecular conformations \cite{LathaConfor}.

% Here- introduce N state model and SF
In this paper we extend the AH model, and allow the electron on the dot to interact with
an $N$-state ``impurity", rather than with a strictly harmonic mode.
Particularly, we introduce the ``spin fermion" (SF) model
with a two-state impurity, e.g. a magnetic spin, see Fig. \ref{Scheme}.
The AH and the SF models
were treated separately in the literature in the context of molecular electronics,
or in relation to the nonequilibrium Kondo physics.
The purpose of this paper is to provide a direct comparison between the transport characteristics of
these two situations, with very simple modeling.
Our goal is to explore whether nontrivial nonequilibrium many-body effects  predicted to show in the AH model:
Franck-Condon blockade and current hysteresis and bistability,
persist when the dot electron interacts with a different type of a scatterer, e.g.,
a magnetic spin.

% technique
We compare the behavior of the AH and SF models in two limits.
First, at high temperatures we use a simple-transparent rate equation method
\cite{Mitra,KochFCBlong,KochThermoE}.
Applying a general small-polaron-type transformation, we reduce
the $N$-state impurity model Hamiltonian into a form suitable for a strong-coupling electron-impurity treatment.
We then study the current-voltage characteristics of the AH and the SF models in the sequential-tunneling limit,
and explore current blockade mechanisms.
We confirm that in the AH model the Franck-Condon blockade (FCB) effect dominates at strong coupling
\cite{KochFCB, KochFCBlong}, but we find that in the SF model this type of blockade is missing altogether.
In the second part of the paper we briefly compare the behavior of the two models in the quantum regime,
in the complementary adiabatic limit (fast electrons and a slow impurity). Particularly, we examine the
possible existence of bistability and hysteresis in the SF model. In this limit we find that
the transport characteristics of the SF and AH models directly correspond, but
that such nonlinear effects, predicted to show up for the AH model, are missing in the SF case.

% this paper
The paper is organized as follows: In Sec. \ref{Model}, we introduce
the general model Hamiltonian and the two examples:
the AH (Sec. \ref{ModelI}) and the SF models (Sec. \ref{ModelII}).
We also discuss these models in the broader
context of transport in a tight-binding network (Appendix).
In Sec. \ref{method} we study the current-voltage characteristics in
the  nonadiabatic limit.
We review the master equation methodology in Sec. \ref{EOM}, and discuss the case with
dissipation in Sec. \ref{equilib}. Numerical results are presented
in Sec. \ref{result}. % examining the current blockade  behavior in the two models.
In Sec. \ref{bis} we discuss the complementary quantum-adiabatic regime of strong electron-metal coupling
and a slow impurity.
Sec. \ref{summary} concludes.
For simplicity, we set $\hbar= 1$, $k_B=1$ (Boltzmann constant), and $e=1$ throughout the paper.

%===========================
\section{Model}
\label{Model}

\subsection{$N$-state impurity}
\label{ModelN}

Our simple modeling of a molecular junction consists a single spin-degenerate molecular electronic level (dot)
of energy $\epsilon_d$. The dot is tunnel-coupled to two voltage-biased metallic contacts.
In the standard Anderson-Holstein model electrons on the dot interact with
equilibrated or unequilibrated harmonic vibrational modes.
We generalize this setup and allow the electron to interact with an $N$-state unit: spin qubit ($N=2$),
large spin ($N>2$), harmonic oscillator ($N\rightarrow \infty$)
or an anharmonic mode with a finite number of bound states.
We refer below to this $N$-state entity as an ``impurity",
and incorporate it in the system-molecular Hamiltonian $H_S$.
The total Hamiltonian comprises the following terms
\bea
H=H_S+H_B+H_{SB}.
\label{eq:H}
\eea
The system Hamiltonian includes the molecular electronic level
(creation operator $d^{\dagger}$), the $N$-state impurity,
and the dot-impurity interaction,
\bea
H_S= \epsilon_d \hat n_d +
\sum_{q=0}^{N-1}\epsilon_q|q\rangle \langle q|
+
\alpha \hat n_d\sum_{q,q'}F_{q,q'}|q\rangle \langle q'|.
\label{eq:HS}
\eea
Here $\hat n_d=d^{\dagger}d$ denotes the occupation number operator for the dot.
The impurity Hamiltonian is written in the energy representation
with the (possibly many-body) states $|q\rangle$, $q,q'=0,1,...,N-1$.
It is coupled to the electron via its operator $F$ with matrix elements $F_{q,q'}$,
$\alpha$ is a dimensionless parameter.
The bath includes two conductors ($\nu=L,R$) comprising noninteracting fermions with
creation (annihilation) operators $a_{\nu,k}^{\dagger}$ ($a_{\nu,k}$),
\bea
H_B=\sum_{\nu,k}\epsilon_k a_{\nu,k}^{\dagger}a_{\nu,k}.
\label{eq:HB}
\eea
The system-bath coupling includes the tunneling Hamiltonian,
\bea
H_{SB}=\sum_{\nu,k}\left(v_{\nu,k}a_{\nu,k}^{\dagger}d +v_{\nu,k}^*d^{\dagger}a_{\nu,k}\right),
\label{eq:HSB}
\eea
with $v_{\nu,k}$ as the tunneling element, introducing the hybridization energy
\bea
\Gamma_{\nu}(\epsilon)=2\pi\sum_{k}|v_{\nu,k}|^2 \delta(\epsilon-\epsilon_k).
\label{eq:Gamma}
\eea
The Hamiltonian (\ref{eq:H})-(\ref{eq:HSB}) can be transformed into a form more
suitable for a perturbative expansion in the tunneling matrix element by means of a unitary-shift transformation.
%To that order, the interaction of the dot with the impurity can be treated exactly.
It is useful to define the impurity Hamiltonian, $H_{imp}=H_S(\hat n_d=1)$, or explicitly
\bea
H_{imp}=\sum_q\epsilon_q|q\rangle \langle q|
+\alpha\sum_{q,q'}F_{q,q'}|q\rangle\langle q'|.
\label{eq:Himp}
\eea
This operator is hermitian and it can be diagonalized with a unitary transformation
\bea
\bar H_{imp}= e^AH_{imp} e^{-A},
\label{eq:U}
\eea
where $A^{\dagger}=-A$ is an anti-hermitian operator in the Hilbert space of the $N$-state impurity.
We now introduce a related unitary operator, $V\equiv e^{A\hat n_d}$. Note that
 $e^{A\hat n_d}de^{-A\hat n_d}=de^{-A}$ and  $e^{A\hat n_d}d^{\dagger}e^{-A\hat n_d}=d^{\dagger}e^{A}$.
Thus, operating on the original Hamiltonian, $\bar H=VHV^{\dagger}$, we reach
\bea
\bar H&=&\sum_{\nu, k}\epsilon_ka_{\nu,k}^{\dagger}a_{\nu,k} + \epsilon_d \hat n_d
\nonumber\\
&+&
\sum_{\nu,k}\left(v_{\nu,k}a_{\nu,k}^{\dagger}de^{-A} +v_{\nu,k}^*d^{\dagger}a_{\nu,k}e^{A}\right)
\nonumber\\
&+& (1-\hat n_d)\sum_q\epsilon_{q}|q\rangle\langle q| + \hat n_d \bar H_{imp}.
%e^{A} H_{imp}e^{-A}.
\label{eq:barH}
\eea
%
%This form is  convenient for developing quantum Master equations approaches, perturbative in the metal-molecule
%tunneling element, but exact, to that order, in the impurity-electron coupling \cite{KochFCB,KochFCBlong}.
We now exemplify this transformation in two limits.
In the standard AH model the impurity corresponds to a harmonic mode
which is coupled through its displacement to the dot.
In the SF model the impurity includes two states,
and the two-state transition operator is coupled to the dot number operator. %XXX name for this type of coupling?
Furthermore, the transformation can be performed on a tight-binding model
with $M$ electronic sites, where each site is coupled to multiple impurities.
In the Appendix we discuss this extension in the context of exciton transfer in chromophore complexes,
considering an anharmonic environment rather than the common harmonic-bath model
\cite{Ex1,Ex2}.

%-----------------------------------------------------------
\subsection{Case I: Harmonic oscillator}
\label{ModelI}

The AH Hamiltonian follows the generic form (\ref{eq:HS})-(\ref{eq:HSB}),
specified as
\bea
H_{AH}=H_S^{AH}+H_B+H_{SB}.
\label{eq:HAH}
\eea
The excess electron on the dot interacts with
an harmonic mode of frequency $\omega_0$, $\epsilon_q=q\omega_0$,
 $q=0,1,2,...$, sometimes referred to as a ``phonon".
The interaction operator allows excitation and de-excitation processes between neighboring
vibrational states,
\bea
F_{q,q'}=\omega_0\sum_{q,q'}\sqrt q|q\rangle \langle q'|\delta_{q'=q-1} +h.c.
\label{eq:Fppp}
\eea
It is more convenient to work with the creation  and annihilation operators,
$b_0^{\dagger}$ and $b_0$, for a boson mode of frequency $\omega_0$.
The molecular Hamiltonian is given by $H_{S}^{AH}=
\omega_0b_0^{\dagger}b_0+ \alpha\omega_0 \left(b_0^{\dagger}+b_0 \right)\hat n_d$,
and the impurity Hamiltonian
\bea
H_{imp}^{AH}=\omega_0b_0^{\dagger}b_0
+ \alpha\omega_0 \left(b_0^{\dagger}+b_0 \right)
\label{eq:HimpAH}
\eea
can be diagonalized
with the (small-polaron) shift transformation (\ref{eq:U}) \cite{Mahan}. The operator $A$ satisfies
\bea
A=\alpha(b_0^{\dagger}-b_0),
\eea
resulting in
\bea
\bar H_{imp}^{AH}=
%e^{A}H_{imp}^{AH}e^{-A}=
\omega_0b_0^{\dagger}b_0-\alpha^2\omega_0.
\eea
We substitute this expression into Eq. (\ref{eq:barH}),
and immediately obtain the standard result
\bea
\bar{H}_{AH} &=&  e^{A\hat n_d}H_{AH}e^{-A\hat n_d}
\nonumber\\
&=&
\sum_{\nu,k}\epsilon_k a_{\nu,k}^{\dagger}a_{\nu,k}
\nonumber\\
&+& \sum_{\nu,k}\left[v_{\nu,k}a^{\dagger}_{\nu,k}de^{-\alpha(b_0^{\dagger}-b_0)} +v_{\nu,k}^*d^{\dagger}a_{\nu,k}e^{\alpha(b_0^{\dagger}-b_0)} \right]
\nonumber\\
&+& \epsilon_d \hat n_d + \omega_0b_0^{\dagger}b_0 - \alpha^2 \omega_0\hat n_d.
\label{eq:barHAH}
\eea
The interaction of electrons with phonons form the ``polaron":
The single-particle dot energies are renormalized, $\epsilon_d\rightarrow \epsilon_d-\alpha^2\omega_0$,
and the tunneling elements are dressed by the
translational operator $e^{-\alpha (b_0^{\dagger}-b_0)}$,
corresponding to a shift in the equilibrium position of the mode when an electron is residing on the dot.

%===============================================
\subsection{Case II: Two-level system}
\label{ModelII}

In the ``spin-fermion model" the excess electron on the dot
is coupled to a two-state system, referred to as a ``spin".
This model has been explored in previous works, for example in Refs. \cite{SF1,SF2,SF3,SF4},
but focus has been placed on the decoherence and dissipative dynamics of the two-level system,
specifically when interacting with a nonequilibrium environment, voltage-biased leads.
Complementing these studies, here we investigate the transport characteristics of the SF model.
The total Hamiltonian (\ref{eq:HS})-(\ref{eq:HSB}) now reads
\bea
H_{SF}=H_{S}^{SF}+H_B+H_{SB},
\eea
with the molecular part
$H_{S}^{SF}=\frac{\omega_0}{2}\sigma_z + \alpha\omega_0\sigma_x\hat n_d$.
Here, $\sigma_{x,y,z}$ denote the Pauli matrices.
The impurity Hamiltonian is hermitian,
\bea
H_{imp}^{SF}=\frac{\omega_0}{2}\sigma_z + \alpha\omega_0\sigma_x,
\eea
and it can be diagonalized with a unitary transformation (\ref{eq:U}).
The generator of this transformation is
\bea
A = i\lambda \sigma_y, \,\,\,\,\, \lambda = \frac{1}{2}\arctan(2\alpha),
\eea
resulting in
\bea
\bar H_{imp}^{SF} = % e^{A}H_{imp}^{SF}e^{-A}=
\frac{\omega_0}{2}\sigma_z + \frac{\omega_0}{2}
\left(\frac{1-\cos2\lambda}{\cos2\lambda} \right)\sigma_z.
\eea
We substitute this expression into Eq.  (\ref{eq:barH})  and reach
\bea
\nonumber
&\bar{H}_{SF}& =  e^{A\hat n_d}H_{SF}e^{-A\hat n_d}
\nonumber\\
&=&
\sum_{\nu,k}\epsilon_k a_{\nu,k}^{\dagger}a_{\nu,k}
\nonumber\\
&+& \sum_{\nu,k}\left[v_{\nu,k} a^{\dagger}_{\nu,k}de^{-i\lambda\sigma_y} + v_{\nu,k}^*d^{\dagger}a_{\nu,k}e^{i\lambda\sigma_y} \right]
\nonumber\\
&+&
\epsilon_d\hat n_d+\frac{\omega_0}{2}\sigma_z +
\frac{\omega_0}{2}
\left(\frac{1-\cos2\lambda}{\cos2\lambda} \right)\sigma_z\hat n_d.
\label{eq:barHSF}
\eea
%with $\kappa(\lambda) = \frac{1-\cos2\lambda}{\cos2\lambda}$. It is instrumental to note the identities: $\sin2\lambda n_d = n_d\sin2\lambda$ and $\cos2\lambda n_d = n_d\cos2\lambda + (1 + n_d)$.
A related shift transformation has been used in Ref. \cite{Sspin} for  studying
the dynamics of a spin immersed in a spin
bath within the noninteracting blip approximation.

Recall that in the  shifted AH model, Eq. (\ref{eq:barHAH}),
electron-phonon coupling shows up in two (polaronic)
features: the dot-metal tunneling elements are dressed,
 and the single particle (dot) energies are renormalized.
In the SF model (\ref{eq:barHSF}) the tunneling operators are similarly dressed
by the interaction parameter $\lambda$, a nonlinear function of the original dimensionless coupling $\alpha$.
Furthermore, the SF model displays an anharmonic characteristic:
the spin gap (energy bias) depends on the charge state of the dot.

%==========================================

\section{kinetic equations for $\Gamma_{\nu}<\omega_0,T_{\nu}$}
\label{method}

In this section we study
the current-voltage characteristics of the AH and SF models of Sec. \ref{ModelI} and \ref{ModelII}
in the classical high-temperature limit and weak dot-metal coupling by
using the kinetic rate equation method of Refs. \cite{Mitra, KochFCBlong,Kochvib,KochThermoE}.

\subsection{Unequilibrated impurity}
\label{EOM}
The shifted Hamiltonian, Eq. (\ref{eq:barHAH}) or Eq. (\ref{eq:barHSF}), can be compacted into
the form  $\bar H=H_B+\bar H_{SB}+\bar H_S$;  $\bar H_{SB}$ includes the dressed tunnel Hamiltonian,
$\bar H_S$ constitutes the dot electron and the impurity, the last three terms in
either Eq. (\ref{eq:barHAH}) or (\ref{eq:barHSF}).
The total Hamiltonian is given in a form conductive for a  perturbative expansion in
the electronic tunnel coupling $v_{\nu,k}$, and we now briefly review the derivation
of a quantum Master equation valid to the lowest order in this parameter, while exact, to
that order, in the impurity-electron coupling.
In the absence of the leads the eigenstates of the molecular system satisfy
\bea
\bar H_{S}|n,q\rangle=\epsilon_{n,q}|n,q\rangle,
\eea
where $n=0,1$ denotes the number of electrons on the dot and $q$ identifies the state of the impurity.
In the AH model [Eq. (\ref{eq:barHAH})], $q=0,1,2,...$ counts the number of excited vibrations and
the eigenenergies of $\bar H_{S}$ obey
\bea
\epsilon_{0,q}&=&q\omega_0,
\nonumber\\
\epsilon_{1,q}&=&\epsilon_d-\alpha^2\omega_0+q\omega_0.
\label{eq:eigenAH}
\eea
In the SF model [Eq. (\ref{eq:barHSF})] $q=\pm$ identifies the state of the spin. There are
four possible molecular eigenstates with energies
\bea
\epsilon_{0,q}&=&q\frac{\omega_0}{2}, %\,\,\,\, \epsilon_{0,+}=\frac{\omega_0}{2}
\nonumber\\
\epsilon_{1,q}&=&\epsilon_d+q \frac{\omega_0}{2}(1+\kappa).
\label{eq:eigenSF}
%\sqrt{1+4\alpha^2}, \,\,\,\,
%\epsilon_{1,+}=\epsilon_d+\frac{\omega_0}{2}\sqrt{1+4\alpha^2}
\eea
Here $\kappa=\left(1-\cos 2\lambda\right)/\cos 2\lambda$. Recall that
$\lambda=\frac{1}{2}\arctan(2\alpha)$, with $\alpha$
as the original (dimensionless) electron-impurity interaction parameter.
Simple manipulations provide $\kappa=\sqrt{1+4\alpha^2}-1$, resulting in
$\epsilon_{1,q}=\epsilon_d+q \frac{\omega_0}{2}
\sqrt{1+4\alpha^2}$.

One can rigorously derive kinetic quantum master equations
for the occupation $P_{q}^n$ of the $|n,q\rangle$ state
when the  metal-molecule coupling is weak, $\Gamma_{\nu}<T_{\nu},\omega_0$.
The standard derivation is worked out from the quantum Liouville equation
%by writing it in an integro differential form and
by applying the Born-Markov approximation, assuming
fast electronic relaxation in the metals and slow tunneling dynamics.
The resulting (bath-traced) reduced-density matrix $\rho_S$ obeys \cite{Qdiss,commentshift}
\bea
\dot \rho_S&=& -i {\rm tr_B}[\bar H_{SB}(t),\rho_S(0)\rho_B]
\nonumber\\
&-& {\rm tr_B}\int_0^t d\tau [\bar H_{SB}(t),[\bar H_{SB}(\tau),\rho_S(t)\rho_B]],
\eea
with $\rho_B$ as the initial state of the two baths (metals),
assumed to be given by a factorized form, with each bath prepared in a thermodynamic equilibrium state
at the temperature $\beta_{\nu}^{-1}$ and a chemical potential $\mu_{\nu}$.
The operators are written in the interaction representation
and the trace is performed over the states of both baths.
Applying the second part of the Markov limit, extending the upper limit of integration to infinity,
this differential equation reduces to the Redfield equation \cite{Qdiss}.
It can be furthermore simplified under
the secular approximation, ignoring coherences between molecular eigenstates.
The result is an equation of motion for the diagonal elements of the reduced density matrix,
$P_{q}^n(t)\equiv \langle q, n|\rho_S(t)  |n, q\rangle$   \cite{Mitra,KochFCBlong, KochThermoE},
\bea
\dot{P_{q}^n}(t) = \sum_{n',q'}\left(P_{q'}^{n'}w_{q'\rightarrow q}^{n'\rightarrow n}
- P_{q}^{n}w_{q\rightarrow q'}^{n\rightarrow n'}\right),
\label{eq:dotP}
\eea
with $w_{q\rightarrow q'}^{n\rightarrow n'}$ as the rate constants for the
$|n, q\rangle \rightarrow |n',q'\rangle$ transition.
Processes that maintain the occupation state of the dot do not contribute
in this low order sequential-tunneling scheme.
Furthermore, the rate constants are additive in this expansion,
$w_{q\rightarrow q'}^{n\rightarrow n'}=\sum_{\nu=L,R}w_{q\rightarrow q',\nu}^{n\rightarrow n'}$
with the $\nu$-bath-induced rates satisfying
\bea
w_{q\rightarrow q',\nu}^{0\rightarrow 1}&=&
s(0,1)\Gamma_{\nu}f_{\nu}(\epsilon_{1,q'}-\epsilon_{0,q}) |M_{q,q'}|^2
\nonumber\\
w_{q\rightarrow q',\nu}^{1\rightarrow 0}&=&
s(1,0)\Gamma_{\nu}\left[1-f_{\nu}(\epsilon_{1,q}-\epsilon_{0,q'})\right] |M_{q,q'}|^2.
\label{eq:w}
\eea
While we had omitted the identifier to the spin state of electrons in the original Hamiltonian,
assuming electronic energies are spin degenerate, the transition rates can be
amended to account for the multiplicity of the $n=1$ level, by introducing the
factors $s(0,1)=2$ and $s(1,0)=1$ \cite{KochThermoE}.
The electronic hybridization is defined in Eq. (\ref{eq:Gamma}),
and it is assumed from now on to be energy independent.
The function $f_{\nu}(\epsilon)=[e^{\beta_{\nu}(\epsilon-\mu_{\nu})}+1]^{-1}$
denotes the Fermi-Dirac distribution of the $\nu$ lead.
The matrix elements
\bea
M_{q,q'}=\langle q|e^{-A}|q'\rangle
\label{eq:M}
\eea
develop from the shift operators decorating the tunneling elements in Eq. (\ref{eq:barH}).
In the AH model these are the familiar Franck-Condon (FC) factors
\cite{NitzanB},
\bea
&&M_{q,q'}^{AH}\equiv\langle q |  e^{-\alpha(b_0^{\dagger}-b_0)}|q'\rangle \,\,\,\,\, q,q'=0,1,2...
\\
&&=
sign(q'-q)^{q-q'}\alpha^{q_M-q_m}e^{-\alpha^2/2}\sqrt{\frac{q_m!}{q_M!}} L_{q_m}^{q_M-q_m}(\alpha^2),
\nonumber
\label{eq:MAH}
\eea
with $q_m=\min\{q,q'\}$,  $q_M=\max\{q,q'\}$, and $L_{a}^{b}(x)$ as the generalized Laguerre polynomials.
In the SF model [Eq. (\ref{eq:barHSF})] this matrix elements are given by  ($q=\pm1$)
\bea
M_{q,q'}^{SF}&\equiv& \langle q |  e^{-i\lambda\sigma_y}|q'\rangle,
\\
M_{q,-q}^{SF}&=& -q\sin \lambda,\,\,\,\,
%  M_{1,-1}^{SF}= -\sin \lambda,
%\nonumber\\
M_{q,q}^{SF}= \cos \lambda.
%M_{1,1}^{SF}= \cos \lambda, \,\,\,\,
\nonumber
\label{eq:MSF}
\eea
Recall, $\lambda = \frac{1}{2}\arctan(2\alpha)$.
The electron current at the $\nu$ contact
can be evaluated within the rate equation formalism at the sequential-tunneling limit \cite{Mitra},
\bea
I_{\nu}=\sum_{q,q'}
\left(P_{q}^0 w_{q\rightarrow q',\nu}^{0\rightarrow 1} -  P_q^1w_{q\rightarrow q',\nu}^{1\rightarrow 0}\right).
\label{eq:curr}
\eea
The correct dimensionality is reached by recovering the prefactor $e/\hbar$.
Eq. (\ref{eq:dotP}) can be readily solved in the long time limit enforcing $\dot P_{q}^{n}=0$.
Substituting the resulting
occupations into Eq. (\ref{eq:curr}),
one can confirm that in steady-state $I\equiv I_L=-I_R$. Our numerical results below display only
steady-state properties.
The formalism discussed here accounts only for sequential-tunneling processes, but it can
be extended without much effort to accommodate next-order (co-tunneling)
terms \cite{Kochvib,KochFCBlong}. One can also generalize this approach and calculate current noise
\cite{KochFCB,KochFCBlong} and other high order cumulants through a full counting statistics analysis
\cite{Komnik,Ora}.

%------------------------------------------------------------

\subsection{Thermally-equilibrated or dissipative impurity}
\label{equilib}

Interaction of the molecular junction with other degrees of freedom (DOF),
solvent, secondary vibrations in the case of a
of molecular junction, nuclear spins, the vibrations in the leads, may further influence the electronic current.
We collect these DOF into an ``environment" and assume that it constitutes a secondary effect for electrons
while it directly dissipates the impurity.
We include this secondary environment in two different ways: (i) by enforcing the impurity
to equilibrate with an additional bath of temperature
$T_h=\beta_h^{-1}$, see Eq. (\ref{eq:eq}) below,
or (ii) by explicitly coupling the impurity  to
a large collection of DOF, noninteracting harmonic oscillators or spins.

{\it Equilibrated impurity.}
The impurity is enforced to equilibrate with a heat bath at $T_h=\beta_h^{-1}$
by enforcing the  ansatz \cite{Mitra},
\bea
P_{q}^n=P^n \frac{e^{-\beta_h\epsilon_{0,q}}}{\sum_{q} e^{-\beta_h\epsilon_{0,q}}}.
\label{eq:equilib}
\eea
We place this expression in Eq. (\ref{eq:dotP}), to solve for the corresponding electronic occupations
($P^1=1-P^0$). In steady-state we find
\bea
P^0=\frac{ \sum_{q,q'}e^{-\beta_{h}\epsilon_{0,q'}} \omega_{q'\rightarrow q}^{1\rightarrow0}}
{ \sum_{q,q'} \left(e^{-\beta_{h}\epsilon_{0,q'}} \omega_{q'\rightarrow q}^{1\rightarrow0}
+ e^{-\beta_{h}\epsilon_{0,q}} \omega_{q\rightarrow q'}^{0\rightarrow1} \right)
}.
\label{eq:eq}
\eea
The electronic occupations are substituted back into Eq. (\ref{eq:equilib})
to directly provide the charge current  (\ref{eq:curr}).

{\it Dissipative impurity.}
We augment the AH Hamiltonian (\ref{eq:HAH}) with a heat heat comprising independent DOF, harmonic oscillators
(bosonic operators $b_{j}^{\dagger}$, $b_j$)
bilinearly coupled (interaction energy $\eta_{j}$)
to the  molecular vibration (bosonic operators $b_{0}^{\dagger}$, $b_0$),
\bea
H_{AH}^{diss}&=&
\sum_{\nu,k}\epsilon_k a_{\nu,k}^{\dagger}a_{\nu,k}
+\sum_{\nu,k}\left(v_{\nu,k}a_{\nu,k}^{\dagger}d +v_{\nu,k}^*d^{\dagger}a_{\nu,k}\right)
\nonumber\\
&+&\omega_0b_0^{\dagger}b_0+ \alpha\omega_0 \left(b_0^{\dagger}+b_0 \right)\hat n_d +\epsilon_d \hat n_d
\nonumber\\
&+& \sum_{j}\omega_j b_{j}^{\dagger}b_{j}+
 \left(b_0^{\dagger}+b_0\right)
\sum_{j}\eta_j\left( b_j^{\dagger}+b_j\right)
\label{eq:HAHdiss}
\eea
Employing the small polaron transformation as discussed in Sec. \ref{ModelI},
$\bar{H}_{AH}^{diss}=e^{A\hat n_d}{H}_{AH}^{diss}e^{-A\hat n_d}$ with $A=\alpha(b_0^{\dagger}-b_0)$,
using the relations $e^{A\hat n_d}b_0^{\dagger}e^{-A\hat n_d}=b_0^{\dagger}-\alpha\hat n_d$
and $e^{A\hat n_d}b_0e^{-A\hat n_d}=b_0-\alpha\hat n_d$,
we get
\bea
\nonumber
\bar{H}_{AH}^{diss} % &=&  e^{A\hat n_d}H_{AH}^{diss}e^{-A\hat n_d}
%\nonumber\\
&=&
\sum_{\nu,k}\epsilon_k a_{\nu,k}^{\dagger}a_{\nu,k} +\sum_{j}\omega_j b_{j}^{\dagger}b_{j}
\nonumber\\
&+& \sum_{\nu,k}\left[v_{\nu,k}a^{\dagger}_{\nu,k}de^{-\alpha(b_0^{\dagger}-b_0)} +v_{\nu,k}^*d^{\dagger}a_{\nu,k}e^{\alpha(b_0^{\dagger}-b_0)} \right]
\nonumber\\
&+& \epsilon_d \hat n_d + \omega_0b_0^{\dagger}b_0 - \alpha^2 \hat n_d\omega_0
\nonumber\\
&+&
\left(b_0^{\dagger}+b_0 -2\alpha \hat n_d\right)
\sum_{j}\eta_j\left( b_j^{\dagger}+b_j\right).
\label{eq:barHAHdiss}
\eea
In this form, the dot electron directly interacts with the phonon environment; this effect is small (as expected)
when $\alpha\ll1$. %%% I think the effect doesn't show up because of secular approximation and isn't limited by alpha..

In the same spirit the SF model can be extended to include a thermal environment,
a harmonic bath or a collection of spins.
In the latter case it is written as
\bea
H_{SF}^{diss}&=&
\sum_{\nu,k}\epsilon_k a_{\nu,k}^{\dagger}a_{\nu,k}
+\sum_{\nu,k}\left(v_{\nu,k}a_{\nu,k}^{\dagger}d +v_{\nu,k}^*d^{\dagger}a_{\nu,k}\right)
\nonumber\\
&+&\frac{\omega_0}{2}\sigma_z + \alpha\omega_0\sigma_x\hat n_d +\epsilon_d \hat n_d
\nonumber\\
&+& \sum_{j}\frac{\omega_j}{2}\sigma_z^j + \sigma_x \sum_j\eta_j\sigma_x^j.
\eea
Applying the shift transformation of Sec. \ref{ModelII}, we arrive at the form
\bea
\nonumber
\bar{H}_{SF}^{diss}% &=&  e^{A\hat n_d}H_{AH}^{diss}e^{-A\hat n_d}
%\nonumber\\
&=&
\sum_{\nu,k}\epsilon_k a_{\nu,k}^{\dagger}a_{\nu,k} + \sum_{j}\frac{\omega_j}{2}\sigma_z^j
\nonumber\\
&+&\sum_{\nu,k}\left[v_{\nu,k} a^{\dagger}_{\nu,k}de^{-i\lambda\sigma_y} + v_{\nu,k}^*d^{\dagger}a_{\nu,k}e^{i\lambda\sigma_y} \right]
\nonumber\\
&+&
\epsilon_d\hat n_d+\frac{\omega_0}{2}\sigma_z +
\frac{\omega_0}{2}
\left(\frac{1-\cos2\lambda}{\cos2\lambda} \right)\sigma_z\hat n_d
\nonumber\\
&+&
\left[ \sigma_x \cos (2\lambda \hat n_d) + \sigma_z \sin (2 \lambda \hat n_d)\right] \sum_j\eta_j\sigma_x^j
\label{eq:barHSFdiss}
\eea
The last term has been obtained by using
the relation
\bea
e^{i\lambda\hat n_d\sigma_y}=\cos(\lambda \hat n_d) +i\sigma_y\sin (\lambda \hat n_d).
\eea
It can be simplified with the identities
$\sin(2\lambda \hat n_d) = \hat n_d\sin2\lambda$ and $\cos (2\lambda \hat n_d) = \hat n_d\cos2\lambda + (1 -\hat n_d)$.

The current-voltage characteristics of the dissipative models
can be readily obtained in the sequential-tunneling limit by extending the rate equation treatment
of Sec. \ref{EOM},
to include a weakly-coupled additional environment.
For example, considering the  SF model (\ref{eq:barHSFdiss}),
the rate equation (\ref{eq:dotP}) becomes
($q,q'=\pm$),
\bea
\dot{P_{q}^n}(t) &=& \sum_{n',q'}
\left(P_{q'}^{n'}w_{q'\rightarrow q}^{n'\rightarrow n}
- P_{q}^{n}w_{q\rightarrow q'}^{n\rightarrow n'}\right)
\nonumber\\
&+&   \sum_{q'\neq q} \left(
P_{q'}^{n}k_{q'\rightarrow q}^{n\rightarrow n}
- P_{q}^{n}k_{q\rightarrow q'}^{n\rightarrow n} \right),
\label{eq:dotPdiss}
\eea
with the metal-induced rates $w_{q\rightarrow q'}^{n\rightarrow n'}$
as in Eq. (\ref{eq:w}), and the heat-bath induced rates
\bea
k_{q\rightarrow q'}^{n\rightarrow n}=\Gamma_{h}(\omega_0)n_{S}[(q'-q)\omega_0][1-n+n\cos(2\lambda)].
\eea
Here and in Eq. (\ref{eq:rateH}) below
the spectral density function,
\bea
\Gamma_h(\omega_0)=2\pi\sum_{j}\eta_j^2\delta(\omega_j-\omega_0),
\label{eq:hybB}
\eea
is evaluated at the impurity energy spacing.
To be consistent with the derivation of the kinetic equation (\ref{eq:dotPdiss}),
this interaction energy should be assumed small, $\Gamma_h\ll \alpha\omega_0$.
The spin distribution function $n_{S}(\omega_0)=[e^{\beta_h\omega_0}+1]^{-1}$
obeys the relation
$n_{S}(-\omega_0)= 1-n_S(\omega_0)$.
We could similarly couple the spin impurity to a harmonic heat bath,
modeling a secondary normal mode environment. In this case
the same rate equation holds, but
the nonzero heat-bath induced rates obey
\bea
k_{q\rightarrow q'}^{n\rightarrow n}&=&\Gamma_{h}(\omega_0)n_B[(q'-q)\omega_0],
%\nonumber\\
%k_{q\rightarrow q-1}^{n\rightarrow n}&=&q\Gamma_{h}(\omega_0)n_{B}(-\omega_0).
\label{eq:rateH}
\eea
The Bose-Einstein distribution function $n_B(\omega_0)= [e^{\beta_h\omega_0}-1]^{-1}$ satisfies
$n_{B}(-\omega_0)= n_B(\omega_0)+1$.
The current [Eq. (\ref{eq:curr})] is computed from the long time solution of  Eq. (\ref{eq:dotPdiss}).

%---------------------------

\subsection{Results}
\label{result}

We study the behavior of the junction in the steady-state limit, %(current, occupation of impurity)
and compare the current-voltage characteristics of the AH and SF models. Particularly,
we wish to understand mechanisms of current suppression in these junctions.
Unless otherwise stated, we used  $\Gamma\equiv\Gamma_L=\Gamma_R$,
$\beta_{L}=\beta_R=20$, $\omega_0=1$. The voltage bias is applied symmetrically,
$\mu_L=-\mu_R$, defining $\Delta \mu=\mu_L-\mu_R$.
The current is given in units of $\Gamma$; the voltage bias $\Delta \mu$, $\Gamma_h$ and $T_{\nu}$, $T_h$
are given in multiples of $\omega_0$.

%=====================================================================

\subsubsection{Molecular eigenenergies and overlap integral}
%\label{FCB}

\begin{figure}[bp]
\vspace{0mm} {\hbox{\epsfxsize=85mm \hspace{-2mm}\epsffile{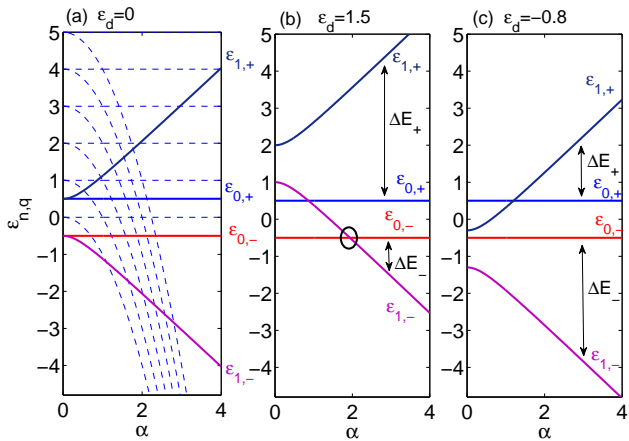}}}
\caption{Eigenenergies  $\epsilon_{n,q}$ of  the
SF model (full) when
(a) $\epsilon_d=0$, (b) $\epsilon_d=1.5$, and (c) $\epsilon_d=-0.8$.
In panel (a) we also display low-lying  ($q=0,1,...,5$)
eigenenergies of the AH molecular Hamiltonian (dashed).
%The degeneracy point $\epsilon_{1,-}=\epsilon_{0,-}$ is encircled in the right panel.
}
\label{FigE}
\end{figure}

We present in Fig. \ref{FigE} the eigenenergies of the
molecular eigenstates $|n, q\rangle$, Eqs. (\ref{eq:eigenAH}) and (\ref{eq:eigenSF}).
%the last three terms in either the AH model, Eq. (\ref{eq:barHAH}) or the SF model, Eq. (\ref{eq:barHSF}).
For simplicity, we include only six levels for the harmonic oscillator.
The energies which do not develop with $\alpha$ correspond to an empty dot, $n=0$.
When an electron is residing on the molecule, the eigenenergies of the two models
show marked qualitative differences:
In the AH model energy spacings between adjacent levels are fixed, $\epsilon_{n,q}-\epsilon_{n,q-1}= \omega_0$,
and the levels bend in a quadratic manner, see Eq. (\ref{eq:eigenAH}).
In contrast, in the SF model the pair with $n=1$  depart;
at small $\alpha$ the departure is quadratic, $\epsilon_{1,+}-\epsilon_{1,-}
\sim \alpha^2\omega_0$, while for large coupling the gap grows linearly with $\alpha$.
In Fig. \ref{FigE} We display results using different gate voltages, $\epsilon_d$,
 to assist us in explaining transport features below.

\begin{figure}[htbp]
\vspace{0mm} {\hbox{\epsfxsize=65mm \hspace{0mm}\epsffile{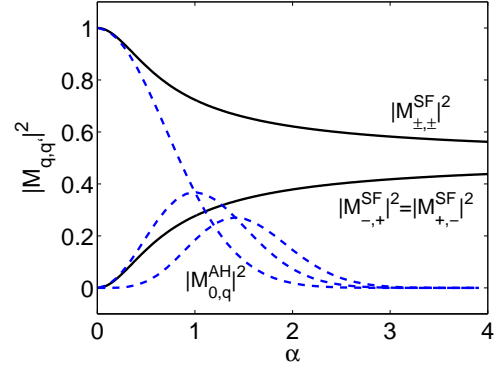}}}
\caption{Dressing elements $|M_{q,q'}|^2$ in the
AH model following
Eq. (\ref{eq:MAH})  with $q=0$ and $q'=0,1,2$ (dashed lines, left to right), and in the
SF model following Eq. (\ref{eq:MSF}),  $q,q'=\pm 1$ (full).
$\omega_0=1$.
}
\label{FigM}
\end{figure}

\begin{figure}[htbp]
\vspace{0mm} {\hbox{\epsfxsize=77mm \hspace{0mm}\epsffile{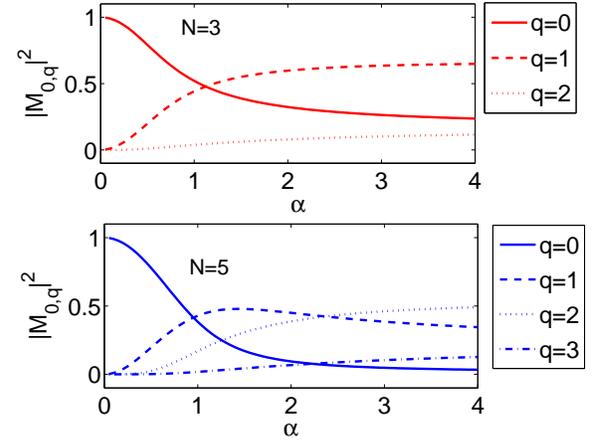}}}
\caption{Dressing elements
$|M_{q,q'}|^2$ for truncated harmonic impurities of $N=3$ and $N=5$ states
with $F_{q,q'}$ from  Eq. (\ref{eq:Fppp}).
}
\label{FigM2}
\end{figure}

The dressing elements of the tunneling Hamiltonian are displayed in Fig. \ref{FigM}.
In the  AH model (dashed lines)
$\langle q |e^{-\alpha(b_0^{\dagger}-b_0)}|0\rangle$ are the common Franck-Condon (FC) factors,
overlap integrals between
the ground vibronic state and excited vibronic levels.
We can interpret the dressing terms of the SF model (full lines) by considering, for example,
the element  $\langle \pm |e^{-i\lambda \sigma_y}|+\rangle$.
Note that when $\alpha\rightarrow \infty$,  %the overlap integral approaches a constant value
$\lambda\rightarrow \pi/4$ and $|\sin(\lambda)|^2=|\cos \lambda|^2=1/2$.
The spin-up state can thus be rotated by an angle $\lambda\leq \pi/4$ to produce
\bea
e^{-i\lambda \sigma_y}|+\rangle = \cos \lambda |+\rangle + \sin \lambda |-\rangle.
\eea
We then overlap the shifted state with the two possible spin outcomes.
We learn from Fig. \ref{FigM} that while in the AH model the FC factors
favor high energy transitions at large $\alpha$, to
realize the Franck-Condon blockade physics, in the SF model this effect is missing and transitions
which do not involve a spin-flip are
favored for all $\alpha$. % ($\pm\rightarrow \pm$);
What about other nanojunctions, with $N>2$ impurities? In Fig. \ref{FigM2} we consider truncated (finite $N$)
harmonic impurities satisfying Eqs. (\ref{eq:Himp})
and (\ref{eq:Fppp}). We display the matrix elements
$M_{0,q}$ obtained from Eq. (\ref{eq:M}), where $e^A$ is the unitary transformation diagonalizing
the relevant impurity Hamiltonian.
We find that already for $N=3$ off-diagonal transitions are favored at large
$\alpha$, once the curves cross and $|M_{0,0}|^2< |M_{0,1}|^2$.
We have also verified (not shown) that for large $N$ we recover the standard FC elements.
%It should be emphasized that the SF model is not unique, and other $N>2$ models may miss the FCB physics
%if the matrix $F$ is designed such that $|M_{0,0}|^2$ is the largest overlap integral for all relevant $\alpha$.
%---------------------------

% eigenenergies

% FC

% FC N>2

%------------------------------------------------

%==========================================================

\subsubsection{Mechanisms of current blockade}
\label{mechanism}

Current blockade, suppression of electronic current
for voltage biases below a certain critical value, may develop through different mechanisms:
(i) In noninteracting models or for weakly-interacting cases
the tunneling current is suppressed in off-resonance situations.
% when the junction is gated such that
%$\epsilon_d$ is placed outside the bias window.
%
We now elaborate on this trivial suppression, then  clarify the related many-body case.
Ignoring interactions, the AH and SF models reduce to the resonant-level model.
The steady-state current can now be calculated exactly, and this Landauer
expression can be expanded in orders of $\Gamma_{\nu}/T_{\nu}$
to provide the lowest order sequential-tunneling limit
\bea
I= \frac{\Gamma_L\Gamma_R}{\Gamma_L+\Gamma_R}\left[f_L(\epsilon_d)-f_R(\epsilon_d)\right].
\eea
If the resonant level, energy $\epsilon_d$, is placed outside
the bias window, an ``off-resonance blockade" (ORB) (current suppression) shows.
At positive bias the blockade is lifted at the critical
voltage $\Delta\mu_c$ satisfying (the Fermi energy is set to zero),
\bea
\Delta \mu_c=2|\epsilon_d|.
\eea
In strongly interacting systems
this off-resonance condition is modified by the many-body interaction parameter $\alpha$.
In general terms, the blockade is lifted when the applied bias is large so as incoming electrons can provide
sufficient energy for making (allowed) transitions between many-body states,
within the relevant order of perturbation theory,
\bea
\Delta\mu_c=2\Delta E, \,\,\,\,\
\Delta E\equiv min|\epsilon_{1,q}-\epsilon_{0,q'}|.
\label{eq:mingap}
\eea
We refer below to this many-body extension of the ORB as
the ``many-body off-resonance blockade" (MB-ORB).
One should note that this effect takes place in both the SF and the AH models.

At low temperatures $T_h/\omega_0\ll1$ only the ground state of the impurity is significantly occupied.
The blockade is then practically determined by a pair of states which are thermally occupied, not necessarily of
the smallest frequency (\ref{eq:mingap}). For example, in the SF model the relevant low temperature energy difference is given by
\bea
\Delta E_{-}&\equiv&|\epsilon_{1,-}-\epsilon_{0,-}|
\nonumber\\
&=&|\epsilon_d - \frac{\omega_0}{2}(\sqrt{1+4\alpha^2}-1)|.
\label{eq:dEm}
\eea
%
%In what follows we demonstrate that the Franck-Condon blockade physics is missing in the SF model.
%However, at strong electron-impurity coupling nontrivial off-resonance blockade (ORB) governs
%the transport behavior,
%
Thermal effects may open up new channels, dramatically reducing the critical voltage:
At high temperatures
both spin states are occupied, thus three other
transitions contribute to the current:
This includes
the transition involving the states $|1,+\rangle$ and $|0,+\rangle$, of spacing
\bea
\Delta E_{+}&\equiv&|\epsilon_{1,+}-\epsilon_{0,+}|
\nonumber\\
&=&|\epsilon_d +\frac{\omega_0}{2}(\sqrt{1+4\alpha^2}-1)|,
\label{eq:dEp}
\eea
and transitions which require a spin-flip ($f$),
\bea
\Delta E_{\pm}^f&\equiv&|\epsilon_{1,\pm}-\epsilon_{0,\mp}|
\nonumber\\
&=&|\epsilon_d \pm\frac{\omega_0}{2}(\sqrt{1+4\alpha^2}+1)|.
\label{eq:dEpf}
\eea
If $\epsilon_d<0$ and $\alpha$ is taken sufficiently large, $\Delta E_{+}$ becomes
 the smallest transition frequency, see Fig. \ref{FigE}(c).
Thus, at {\it negative gating} the blockade region contracts from $\Delta E_{-}$ to $\Delta E_{+}$
when we increase the temperature from $T_h/\omega_0\ll1$ to $T_h/\omega_0\sim 1$.
This strong effect is displayed below in
 Fig. \ref{Figdiss2}.

(ii) The ``Franck-Condon blockade" effect
%describes the suppression of current
dominates the AH physics at  strong electron-phonon coupling \cite{KochFCB,KochFCBlong}.
This is because at large shifts $\alpha\gg1$ transitions from $q=0$ to high vibronic states ($q'>>q$) are favored
over low-lying states, see the structure of the FC factors in Eq. (\ref{eq:MAH}).
Thus, the (low-bias) current is suppressed and
the blockade is lifted only at large bias
once incoming electrons have sufficient energy to excite high vibronic states.

(iii) Repulsion (strength $U$) between electrons on the dot may drive
the ``Coulomb blockade" effect if $\Gamma_{\nu}<T_{\nu}$ and $U>\Gamma_{\nu}$. %XXX
We do not consider this type of Blockade in the present analysis though extensions are immediate
\cite{KochThermoE}.

In what follows we exemplify
current suppression in the AH and SF models.
Recall that the Franck-Condon blockade physics is missing in the SF setup since its overlap
matrix elements (\ref{eq:MSF}) do not cross. As a result,
at weak electron-impurity coupling the transport behavior in the two models is expected to be
similar, controlled
by the ORB. At intermediate coupling (when the FC factors obey $|M_{0,0}|>|M_{0,1}|>|M_{0,2}|...$)
both models are affected by the MB-ORB, renormalizing the suppression region.
At strong coupling the AH model is controlled by the FC factors,
while the behavior of the SF model is determined by the  MB-ORB physics.

%------------------------------
% I-V ed=0
\begin{figure}[htbp]
\vspace{0mm}
{\hbox{\epsfxsize=65mm \hspace{2mm}
\epsffile{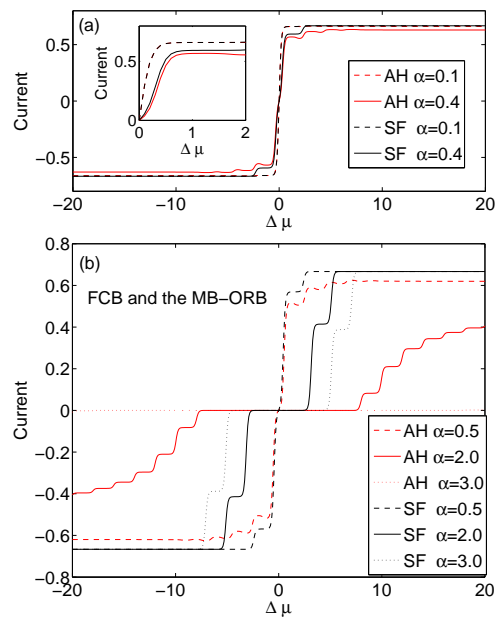}}}
\caption{
Current-voltage characteristics of the AH and SF models at $\epsilon_d=0$
with weak (a) and strong (b) electron-impurity coupling.
(a) The inset zooms on weak coupling features, demonstrating
the similarity, and onset of deviations, between the models, as coupling increases.}
%\omega_0=1$, $\beta_{\nu} = 20$.}  %20 levels for HO
\label{FigJ1}
\end{figure}

% I-V ed>0
\begin{figure}[htbp]
\vspace{0mm}
{\hbox{\epsfxsize=70mm \hspace{2mm}
\epsffile{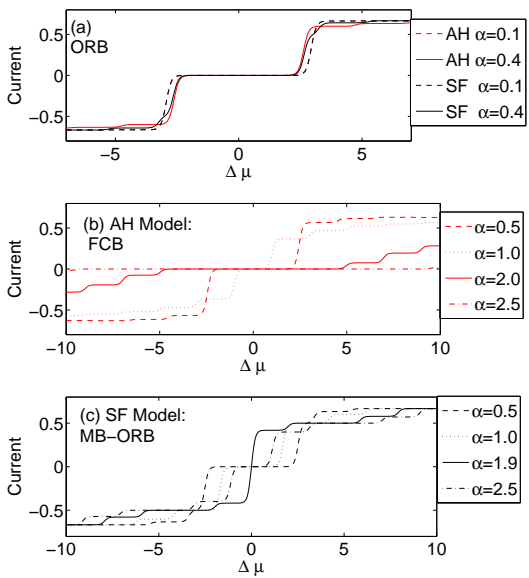}}}
\caption{Current-voltage characteristics in a gated  $\epsilon_d=1.5$ junction
at weak (a) and strong  (b-c) coupling.
Different types of blockade play a role:
(a) ORB at weak interactions,
(b) FCB in the AH model, and
(c) MB-ORB in the strongly-interacting SF model.
 }  %20 levels for HO
\label{FigJ2}
\end{figure}
%------------------------------

\subsubsection{unequilibrated impurity}
\label{FCB}

We display the current-voltage characteristics of the AH and the SF junctions
in Figs. \ref{FigJ1}-\ref{FigJ2}.
The dot energy is placed either at the center of the bias window,
$\epsilon_d=0$, or, under positive gating conditions we set $\epsilon_d=1.5$ \cite{comment1}.
In the weak coupling limit ($\alpha\ll1$) both models show similar features, particularly,
an off-resonance suppression of the tunneling current, see Fig. \ref{FigJ2}(a).
At strong coupling $\alpha\sim 2$,
the models show current blockade, however the underlying cause
differs. In  the AH model the current is suppressed due to the behavior of the FC
factors, favoring distant-energetic vibronic transitions;
in the SF model diagonal, $q\rightarrow q$, transitions always dominate.
Instead, the current is suppressed  by the MB-ORB effect:
As we increase the coupling to the impurity, the molecular frequency
relevant for the onset of current develops as
%(the state $|0,-\rangle$ is initially occupied)
$\Delta E_- =|\epsilon_d - \frac{\omega_0}{2}(\sqrt{1+4\alpha^2}-1)|$.
When $\epsilon_d=0$ the blockade region is monotonically increasing
with $\alpha$, in a linear fashion for large $\alpha$.
In the gated $\epsilon_d>0$ case
the blockade physics is more involved;
the current is suppressed at sufficiently low biases if
the bare energy $\epsilon_d$ is tuned {\it away} from the special point of degeneracy $\Delta E_-=0$, satisfying
\bea
\epsilon_d = \frac{\omega_0}{2}\left(\sqrt{1+4\alpha^2}-1\right).
\label{eq:res}
\eea
This point
is encircled in Fig. \ref{FigE}(b), taking place at $\alpha=1.94$ for $\epsilon_d=1.5$.
Fig. \ref{FigJ2}(c) shows that
the low-bias current is indeed suppressed in the SF model when $\alpha\neq 1.9$.
%
%Note that the earlier crossing of  $\epsilon_{1,-}$ with $\epsilon_{0,+}$  (see
%Fig. \ref{FigE},  $\alpha=0.8$)
%involves  spin flip, but given the small overlap matrix element involved, this process does not dominate
%the current.

%Once the voltage bypasses the relevant energy gap $\Delta \mu>E_b$,
%the current rapidly increases to its full value.
Note that the MB-ORB effect takes place in the AH model as well:
Besides the FC physics, the off-resonance blockade is lifted at level crossings when
$\epsilon_{1,q}=\epsilon_{0,q}$, or
$\epsilon_d=\alpha^2\omega_0$, see  the $\alpha=0.5, 1$ data lines in
Fig. \ref{FigJ2}(b).
However, at large coupling ($\alpha>1$) the MB-ORB effect is marginal in the AH model,
and the FCB physics dominates.

\begin{figure}[htbp]
\vspace{0mm}
{\hbox{\epsfxsize=80mm
\epsffile{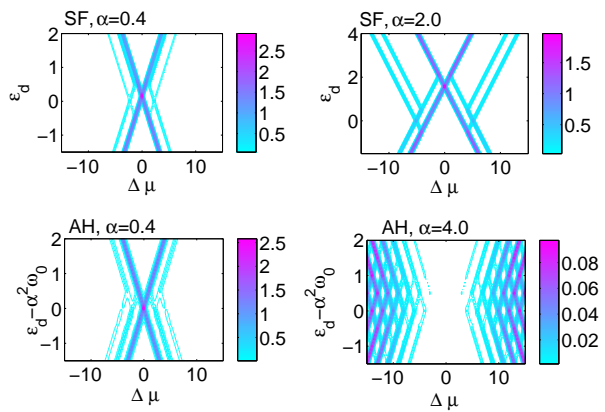}}}
\caption{Differential conductance plots of the SF (top) and the  AH (bottom) models
as a function of gate ($\epsilon_d$) and applied bias voltage $\Delta \mu$
at weak  and strong coupling, as indicated in the figure.
}  %20 levels for HO
\label{FigJ3}
\end{figure}

\begin{figure}[h]
\vspace{0mm}
{\hbox{\epsfxsize=75mm
\epsffile{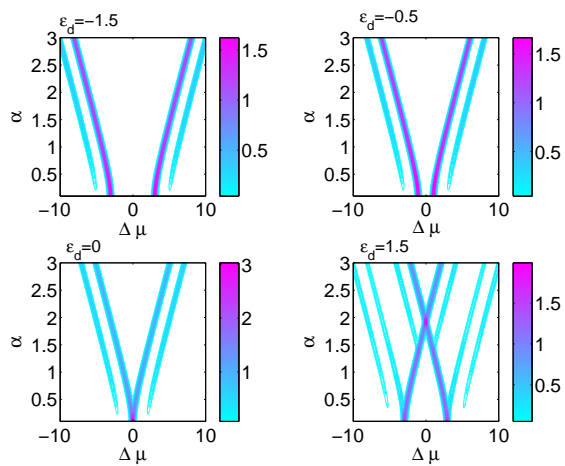}}}
\caption{Differential conductance plots of the SF model
as a function of electron-spin interaction ($\alpha$) and the bias voltage $\Delta \mu$
at different gating, as indicated in the figure.
}  %20 levels for HO
\label{FigJ4}
\end{figure}

Conductance  plots  ($dI/d\Delta \mu$) are presented in Fig. \ref{FigJ3}. The
AH model demonstrates the FCB physics, the development of the gap with increasing $\alpha$.
The SF model shows uneven level spacings, the result of molecular anharmonicity,
and the development of the MB-ORB effect
away from the degeneracy point at $\Delta E_-=0$. % at $\epsilon_d=1.56$ ($\alpha=2$).
% DDD
In Fig. \ref{FigJ4} we complement this analysis and present
the low-temperature conductance as a function of bias voltage and electron-impurity interaction parameter $\alpha$.
We find that at negative gating %(the Fermi energy is set at zero)
the blockade region monotonically
 increases with $\alpha$.
For positive gating there is a particular solution of Eq. (\ref{eq:res}),
resulting in a resonance behavior.
%------------------------------

% dI/dV

% dI/dV

%-------------------------------------------

%======================================================

\subsubsection{Energy Dissipation and thermal equilibration}

The behavior of the AH model with an equilibrated vibration was considered in several studies, see Refs.
\cite{Mitra,KochFCB,KochFCBlong}. Models with explicit secondary heat baths were reviewed in Ref.
\cite{GF1}. In the context of
the Franck-Condon blockade physics
it was shown (in the sequential-tunneling regime)
that the blockade becomes {\it more rigorous} when the harmonic mode is equilibrated;
when the mode is
unequilibrated tunneling electrons
may leave the molecular system with an excited vibration, and
subsequent tunneling processes can continue and increase the excitation state
 \cite{KochFCB}.
When co-tunneling processes are included,
the AH model with equilibrated vibrations shows a significant increase in current for small biases,
$\Delta \mu<\alpha^2\omega_0$, yet the FCB survives.

In this section we study the role of dissipation effects and equilibration
on the current-voltage characteristics of the SF model.
The role of mode equilibration is explored using the ansatz (\ref{eq:equilib}).
The more gentle introduction of dissipation effects
is studied using Eqs. (\ref{eq:HAHdiss})-(\ref{eq:rateH}).

\begin{figure}[bp]
\vspace{0mm} {\hbox{\epsfxsize=70mm \hspace{4mm}\epsffile{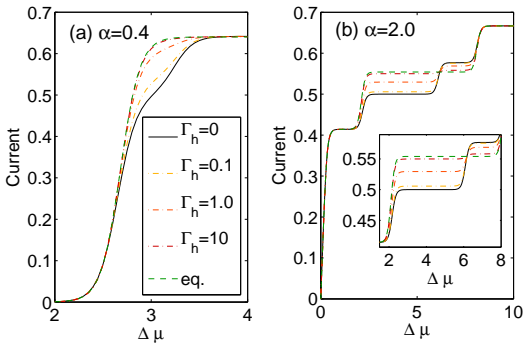}}}
\caption{
Mild effect of mode equilibration on the current at $\epsilon_d>0$ in the
(a) weak coupling limit, and
(b) at strong coupling;
the inset zooms on the region of interest.
The legend describes all panels:
(full) excluding a heat bath,
(dashed-dotted lines) including a dissipative spin bath at different couplings,
and (dashed)
once enforcing impurity equilibration as in Eq. (\ref{eq:equilib}).
We used $\epsilon_d=1.5$ and $T_h=0.05$.
}
\label{Figdiss}
\end{figure}

We found that the equilibration of the impurity
did not affect the transport behavior of SF junctions when $\epsilon_d=0$
(not shown).
In Fig. \ref{Figdiss} we thus display the current at positive gating, $\epsilon_d=1.5$.
%In this regime the first encountered molecular frequency corresponds to $g\rightarrow e$ transitions.
%The only populated state in the absence of dissipation for small bias is $\ket{0,-}$,
%and therefore, as expected, the suppression region
%is unaffected by the enforced equilibration of the impurity,
%and the current is only enhanced after the first step.
% Dvira The current here again should increase at high T, when the transition from e0+ to e1- is allowed.
First, we confirm that the dissipative model interpolates correctly between the
 isolated case $\Gamma_h=0$ and the equilibrated
$\Gamma_h/\omega_0>\alpha$ limit. The latter choice of parameters goes beyond the weak (heat bath-impurity)
coupling assumption underlying the derivation of Eq. (\ref{eq:dotPdiss}).
It is included here for demonstrating
that the dissipative model provides seemingly meaningful results even at strong dissipation $\Gamma_h$.
It is interesting to note that coupling to a secondary bath may increase the current, compared to the case
without this bath, or decrease it, see panel (b) in Fig. \ref{Figdiss}.

\begin{figure}[htbp]
\vspace{0mm} {\hbox{\epsfxsize=70mm \hspace{4mm}\epsffile{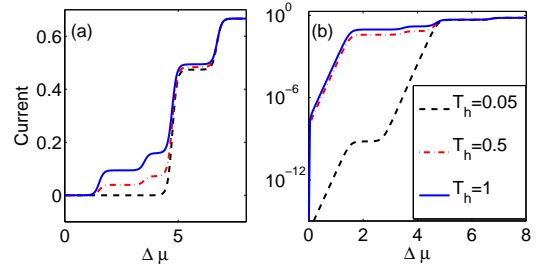}}}
\caption{
Strong influence of mode equilibration on the current at $\epsilon_d<0$
(a) linear scale, (b) logarithmic scale, displaying steps at low temperatures.
%Current-voltage characteristics of the SF model with an equilibrated mode, $\epsilon_d=-0.8$,
The temperature of the electronic baths is (as before) $T_{\nu} = 0.05$. $T_{h}$ is indicated in the figure,
and we used $\alpha = 2$ and $\epsilon_d=-0.8$.
}
\label{Figdiss2}
\end{figure}

Thermal effects influence the current only modestly at positive gating as observed in
Fig. \ref{Figdiss},
particularly leaving intact the MB-ORB region.
This is true as long as $\Delta E_-$ is the smallest allowed transition frequency,
see Fig. \ref{FigE}(b).
%when occupying spin-up states does not
%modify the critical voltage $\Delta\mu_c$.
In contrast, at negative gating ($\epsilon_d<0$)
 dissipation or an enforced equilibration
markedly influence the current, contracting
the blockade region, see Fig. \ref{Figdiss2}. As discussed below Eq. (\ref{eq:dEp}),
this is because $\Delta E_+$ is the smallest molecular frequency
at negative gating and large $\alpha$,
see  Fig. \ref{FigE}(c).
Therefore, by thermally-occupying spin-up states we cut-down the critical voltage $\Delta\mu_c$ from
$\Delta E_-$ to $\Delta E_+$, further
exposing the other $\Delta E_{\pm}^f$ transitions as steps
in the current-voltage characteristics.

%--------------------------------------
% IV with dissipation


%--------------------------------------
\section{Adiabatic limit $\Gamma_{\nu}>\omega_0$}
\label{bis}
%: Bistability and hysteresis

% parameters
In the previous section we studied the nonadiabatic high temperature limit, $\Gamma_{\nu}<\omega_0, T_{\nu}$,
while allowing the electron-impurity interaction energy to become arbitrary large.
In this section we focus on the opposite adiabatic regime
of large tunneling elements $\Gamma_{\nu}>\omega_0$, small $\alpha$, %relatively small coupling to the impurity,
and low temperatures $T_{\nu}<\omega_0$.

% hysteresis
The possible existence of more than one steady-state in molecular junctions,
and potential mechanisms of bistability, switching, and hysteresis,
have been topics of interest and controversy in the past decade.
%\cite{NitzanBi,NitzanBiGF,RabaniBi1,RabanBi2,Brat1,Brat2,Kosov,Lothar}.
While early considerations adopted
the Born-Oppenheimer mean-field approximation \cite{NitzanBi,Kosov}
and perturbative treatments \cite{NitzanBiGF},
more recent studies addressed this problem using brute-force numerically exact simulation
tools \cite{Lothar,RabaniBi1,RabaniBi2}.

% here
In this section we consider  the existence of
bistability, hysteresis and switching in molecular junctions consisting an anharmonic impurity,
the SF model.
These effects, discussed so far in detail within the AH model,
are in principle not limited to strictly harmonic impurities.
%However, here we show that
%the SF model cannot recover these nontrivial functionalities within relevant-consistent parameters.
%
Our analysis goes back to the simple mean-field treatment of Galperin et al. \cite{NitzanBi}
valid in the limit of a large tunneling element $\Gamma_{\nu}>\omega_0$.
This mean-field approach naturally fails in certain physical regimes \cite{Brat1,Brat2}, yet
it serves as a valid starting point for comparing the AH and SF models, for considering phenomenology
preceding extensive numerical treatments \cite{RabaniBi1,RabaniBi2}.
%We now apply the Born-Oppenheimer approach of Ref.  \cite{NitzanBi} on the SF model.
We find that the self-consistent equations, for the dot occupation and charge current, have a related form
in the AH and SF models. However,
bistability and hysteresis are missing in the latter case,
considering the allowed-consistent range of parameters.

%The electronic level shift corresponds here to a very slow spin,
%thus the electron on the dot itneracts with a constant
%spin state, itself calculated from the averaged electron occupation on the dot.

We begin by introducing a variant of the dissipative SF model, complementing the models of Sec. \ref{equilib},
\bea
H_{SF}^{diss}&=&
\sum_{\nu,k}\epsilon_k a_{\nu,k}^{\dagger}a_{\nu,k}
+\sum_{\nu,k}\left(v_{\nu,k}a_{\nu,k}^{\dagger}d +v_{\nu,k}^*d^{\dagger}a_{\nu,k}\right)
\nonumber\\
&+&\epsilon_d\hat n_d+\frac{\omega_0}{2}\sigma_z + \alpha\omega_0\sigma_x\hat n_d
\nonumber\\
&+& \sum_{j}\omega_jb_j^{\dagger}b_j  + \sigma_z
\sum_j\eta_j\left(b_j^{\dagger}+b_j\right).
\label{eq:SFH}
\eea
The impurity polarization is coupled to displacements
of harmonic oscillators in a secondary heat bath, itself prepared in a thermodynamic state at temperature $T_h$.
In the adiabatic limit $\Gamma_{\nu}>\omega_0$ tunneling electrons are fast and the two-state impurity
is slow.
Under a Born-Oppenheimer timescale-separation approximation
a dissipative spin Hamiltonian can be defined,
\bea
H_{S}&=&
\frac{\omega_0}{2}\sigma_z + M\sigma_x n_d
\nonumber\\
&+& \sum_{j}\omega_jb_j^{\dagger}b_j  + \sigma_z
\sum_j\eta_j\left(b_j^{\dagger}+b_j\right),
\label{eq:SFBO}
\eea
consisting slow DOF.
Here $n_d={\rm tr}[\rho \hat n_d]$ stands for the expectation value of the
dot number operator in the steady-state limit;
$\rho$ is the total density matrix.
The definition $M\equiv\alpha\omega_0$ for the electron-spin interaction energy takes us back to the notation of
Ref. \cite{NitzanBi}. % to allow direct comparison between results.
However, while in the AH model the related electron-averaged Hamiltonian includes only
harmonic modes, resulting in
 an exact quantum Langevin equation treatment \cite{NitzanBi},
Eq. (\ref{eq:SFBO}) reduces to the more complex ``spin-boson" Hamiltonian;
by further defining the spin tunneling element as $\Delta\equiv 2M n_d$ we recover the usual form
of this model.

It is useful to define
the spectral density function, $J(\omega)=4\pi\sum_{j}\eta_j^2\delta(\omega-\omega_j)$,
enclosing the interaction of the spin with the boson heat bath. It is
assumed here to take an Ohmic form,
\bea
J(\omega)=2\pi\omega K e^{-\omega/\omega_c},
\eea
with $\omega_c$ as the cutoff frequency of the heat bath and $K$ a dimensionless damping parameter.

The thermodynamic properties and the dynamical behavior of the spin-boson model were explored in
details in different limits \cite{Weiss}.
If the damping is weak ($K\ll1$)
%and at low temperatures $T<\omega_0,\Delta$
it can be shown that the long-time bath-traced coherence obeys in the Ohmic case the expression
%the expression \cite{Weiss}
%
\bea
\langle \sigma_x\rangle\sim-\frac{\Delta_{eff}^2}{\Delta\Omega}\tanh\frac{\Omega}{2T_h},
\label{eq:sig}
\eea
valid beyond the noninteracting blip approximation \cite{Weiss}.
Here $\Omega^2=\Delta_b^2(1+2K\mu)$,
$\Delta_b=[\omega_0^2+\Delta_{eff}^2]^{1/2}$, $\mu=\Re\Psi(i\Delta_b/2\pi T_h)-\ln(\Delta_b/2\pi T_h)$
with $\Re$ denoting the real part of $\psi$, the digamma function.
The effective  tunneling element, between spin states, is given by
$\Delta_{eff}=\Delta\left[ \Gamma(1-2K)\cos (\pi K)\right]^{1/2(1-K)} (\Delta/\omega_c)^{K/(1-K)}$;
$\Gamma$ stands here for the Gamma function \cite{Weiss}.
While we could continue our analysis with this expression,
we simplify it so as to arrive at the expressions of Ref. \cite{NitzanBi}.
We thus consider the limits
$\omega_c\gg \Delta$,
$\omega_0>\Delta$, and $T_h<\Delta$. We can now approximate $\Delta_{eff}\rightarrow \Delta$,
$\Delta_b\rightarrow\omega_0$, reducing  Eq. (\ref{eq:sig}) to
\bea
\langle \sigma_x\rangle&\sim&
-\frac{\Delta}{\omega_0(1+K\mu)},
%\nonumber\\
%= -\frac{2M n_d }{\omega_0(1+K\mu)}.
\label{eq:sigma}
\eea
recall that $\Delta={2M n_d}$.
The denominator describes the renormalization of the spin splitting due to
the coupling to a heat bath.
We now turn our attention to the fast, fermionic, degrees of freedom, and define the Hamiltonian
\bea
H_{F}&\equiv&
\tilde \epsilon_d(n_d)\hat n_d +
\sum_{\nu,k}\epsilon_k a_{\nu,k}^{\dagger}a_{\nu,k}
\nonumber\\
&+&\sum_{\nu,k}\left(v_{\nu,k}a_{\nu,k}^{\dagger}d +v_{\nu,k}^*d^{\dagger}a_{\nu,k}\right),
\label{eq:F}
\eea
with the shifted dot  energy
\bea
\tilde\epsilon_d(n_d)=\epsilon_d-\frac{2M^2n_d}{\omega_0(1+K\mu)}.
\label{eq:edeff}
\eea
The shift is referred to as a ``reorganization energy",
$\epsilon_{reorg}\equiv M^2/[\omega_0(1+K\mu)]$, and it
absorbs the response of the impurity and its attached bath to charge occupation on the dot.
%The electronic level shift corresponds here to a very slow spin,
%thus the electron on the dot itneracts with a constant
%spin state, itself calculated from the averaged electron occupation on the dot.

The electronic Hamiltonian, Equations (\ref{eq:F})-(\ref{eq:edeff}),
is parallel to the result of Galperin et al. \cite{NitzanBi}.
Repeating their arguments,
bistability may, in principle, develop since
the following coupled equations can take more than one solution,
\bea
n_d&=&\frac{\Gamma_L}{\pi(\Gamma_L+\Gamma_R)}
\arctan \left[x+2\frac{\mu_L}{(\Gamma_L+\Gamma_R)}\right]
\nonumber\\
&+&
\frac{\Gamma_R}{\pi(\Gamma_L+\Gamma_R)}\arctan \left[x+2\frac{\mu_R}{(\Gamma_L+\Gamma_R)}\right]
\nonumber\\
&+& \frac{1}{2},
\label{eq:nd1}
\\
n_d&=&\frac{\Gamma_L+\Gamma_R}{4\epsilon_{reorg}}x + \frac{\epsilon_d}{2\epsilon_{reorg}},
\label{eq:nd2}
\eea
The first equation here describes the steady-state zero-temperature
expectation value of the dot occupation under the electronic Hamiltonian (\ref{eq:F}).
The second equation corresponds to
 the shifted dot energy (\ref{eq:edeff}) with
($\mu_F=(\mu_L+\mu_R)/2=0$)
$x\equiv-2\tilde \epsilon_d/(\Gamma_L+\Gamma_R)$.
To treat the case of nonzero temperatures one should retract to Eq. (\ref{eq:sig}) and employ
the finite temperature solution for the dot occupation, replacing Eq. (\ref{eq:nd1}). % XXX

\begin{figure}[htbp]
\vspace{0mm} {\hbox{\epsfxsize=70mm \hspace{4mm}\epsffile{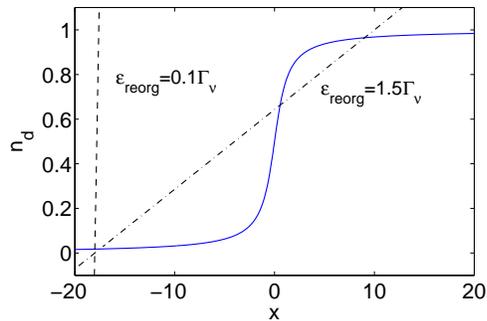}}}
\caption{Electronic dot occupation in the SF model, quantum adiabatic limit,
with $\epsilon_d=4.5$, $\Delta\mu=0$,
and $\Gamma_{\nu}=0.25$.
The full line was  generated from Eq. (\ref{eq:nd1}).
Eq. (\ref{eq:nd2}) provides the dashed (dashed-dotted) lines,
based on data consistent (inconsistent) with the derivation of Eq. (\ref{eq:nd2});
the dashed-dotted line is included here for demonstrating that multiple solutions can show only when
$\epsilon_{reorg}/\Gamma>1$, deviating from the assumptions leading to Eq. (\ref{eq:nd2}).
}
\label{Fbis}
\end{figure}

We now point that in developing Eq. (\ref{eq:edeff}) we have made the assumption $\Delta <\omega_0$,
translating to $\alpha<1$. Given that $\Gamma_{\nu}>\omega_0$, we conclude that our analysis is
valid only when
$\epsilon_{reorg}\sim \alpha^2\omega_0<\Gamma_{\nu}$.
This implies a large slope in Eq. (\ref{eq:nd2}), providing only one solution, see Fig. \ref{Fbis}.
It can be similarly shown that multiple solutions are missing in
the opposite $\Delta>\omega_0$ limit.

Thus, when the electron is coupled to a dissipative
 two-state mode, we reach adiabatic equations which
directly correspond to those obtained in the dissipative AH model.
However, multiple solutions are missing in the  SF model
at the level of the mean-field approximation.
Numerically exact simulations should be performed to reach conclusive results.
%We note that the SF model (\ref{eq:SFH})
%simulate using numerically exact techniques
%in which the impurity spectrum is naturally truncated.
Particularly fitting are
influence functional path integral approaches
in which the impurity spectrum is naturally truncated \cite{INFPIy}.

To complement transport studies,  Sec. \ref{method},
we further write the adiabatic limit of the charge current,
a Landauer expression,
\bea
I=\frac{1}{2\pi}\int d\epsilon\frac{\Gamma_L\Gamma_R[f_L(
\epsilon)-f_R(\epsilon)]}{[\epsilon-\tilde \epsilon_d(n_d)]^2+(\Gamma_L+\Gamma_R)^2/4}.
\eea
The (assumed energy independent) hybridization energy $\Gamma_{\nu}$ was defined in Eq. (\ref{eq:Gamma}).
The system shows an off-resonance blockade, and the critical bias is (simply) linearly
reduced by the reorganization energy,  see Eq. (\ref{eq:edeff}).

%----------------------------
%

%--------------------------------------------------------------

\section{Summary}
\label{summary}

%In this work we examined transport behavior in nanojunctions
%in which electrons on the junction interact
%with a naturally anharmonic local mode.
%
%Our analysis allows for a direct comparison between distinct models, for elucidating the role of
%different scatterers on electron conduction in nanojunctions.

The Anderson-Holstein model provides a minimal
description of molecular junctions, by
including the interaction of electrons in the molecule with a harmonic-vibrational mode.
The spin-fermion model describes simplified nonequilibrium Kondo-like
systems in which conducting electrons interact with a spin impurity.
Our goal here has been to complement studies of the AH model,
and analyze the role of mode anharmonicity on nonlinear transport characteristics,
blockade physics and possible bistability.

In the first - main part of the paper we considered the nonadiabatic (slow electron) limit.
We transformed the AH and the SF models into a comparable form, suitable for
a perturbative expansion in the tunneling element, where to that order, the coupling of the dot electron
to the impurity (vibrational mode or spin) is included to all orders.
In the limit of weak electron-impurity coupling the two models support similar
transport behavior.
At strong electron-impurity interactions significant deviations arise. Principally,
the SF model does not support the analog of the
Franck Condon blockade physics which governs the behavior of the AH model.
However, the SF model does show a nontrivial many-body off-resonance current suppression;
the off-resonance regime is determined by
a nonlinear function of the electron coupling to the impurity, and by the gate voltage ($\epsilon_d\neq0$).
In the second part of the paper we briefly analyzed the adiabatic limit at low temperatures. % XXX
Based on mean-field arguments, we pointed out that
that electron occupation and the charge current in the SF model obey adiabatic equations analogous
to those reached in the AH system.
However, multiple solutions are absent in the case of a two-state impurity, thus
nonlinear transport effects such as bistability and hysteresis are missing, at this level of approximation.

The AH and the SF models discussed in this paper
can describe hybrid physical scenarios beyond molecular junctions \cite{hybrid},
for example,
nanomechanical systems in which the conducting electrons interact with mechanical modes \cite{FCBexp},
and {\it photon} assisted electron transport situations, through quantum dot systems \cite{cavity}.
In future work we will examine
the correspondence in transport behavior
between harmonic and anharmonic-mode models
using numerically exact methodologies \cite{INFPIy}.

%assisted by cavity photons

%perfectly adequate: the harmonic approximation is ubiquitous in optics
%and spectroscopy and may represent an optical cavity exactly,
%while $N$-level systems are popular in the quantum information ($N=2$),
%thermodynamics($N=3$), energy transport in biology($N\ge2$), etc...

%charged state. Further, we considered effects of dissipation
%and found that 1. Hidden channels may be revealed with increasing temperature
%of the heat bath. 2. Non-linear effects, such as NDC may suffer from high
%temperatures of the vibrational "pool" present in such junctions.

%========================
\begin{acknowledgments}
The work of LS was supported by an Early Research Award of DS, by
an Ontario Graduate Scholarship, and by the Jim Guillet Chemistry Graduate Scholarship.
DS acknowledges support of the Discovery Grant Program from
the Natural Sciences and
Engineering Research Council of Canada.
\end{acknowledgments}
%========================

%------------------------------------------------------------------------------------------

%-----------------------------
%=======================
\renewcommand{\theequation}{A\arabic{equation}}
\setcounter{equation}{0}  % reset counter

\section*{Appendix: Collections of harmonic modes or spins}

\label{bath}

The general transformation discussed in Sec. \ref{ModelN} can be performed on an extended model
with $M$ spin-degenerate electronic sites, $m=1,2,...,M$, where each site is coupled to multiple impurities.
In the case of the generalized AH model this constitutes a collection of phonons, and
the tight-binding network is given by
\bea
H_{AH}^M&=& %\sum_{\nu, k}\epsilon_kc_{\nu,k}^{\dagger}c_{\nu,k} +
\sum_{m}\epsilon_m \hat n_m
+\sum_{m,m'}\left(v_{m,m'}a_{m}^{\dagger}a_{m'} +v_{m',m}^*a^{\dagger}_{m'}a_{m}\right)
\nonumber\\
&+&
\sum_{p}\omega_{p}b_{p}^{\dagger}b_{p} +
\sum_m \hat n_m \sum_{p}\alpha_{m,p}\omega_{p}\left(b_{p}^{\dagger}+b_{p} \right).
\label{eq:HAHbath}
\eea
Here $a_m^{\dagger}$ ($a_m$) are creation (annihilation) fermionic operators.
The  set of local phonons (creation operator $b_{p}^{\dagger}$) is coupled to
the electronic number operator of site $m$, $\hat n_m$,  with the dimensionless parameter $\alpha_{m,p}$.
%Besides problems of electron transfer in condensed phases,
%this model has been used to describe excitonic energy transfer in protein environment, see e.g., \cite{Ex1,Ex2}.
The polaron-transformed Hamiltonian, an extension of Eq. (\ref{eq:barHAH}), is given by
\bea
\bar H_{AH}^M&=& %\sum_{\nu, k}\epsilon_kc_{\nu,k}^{\dagger}c_{\nu,k} +
\sum_{m}\left(\epsilon_m-\sum_p \alpha_{m,p}^2\omega_{p}\right)  \hat n_m
\nonumber\\
&+&\sum_{m,m'}\left(v_{m,m'}a_{m}^{\dagger}a_{m'}e^{(A_m-A_{m'})}  +h.c. \right)
%+v_{m+1,m}^*a^{\dagger}_{m+1}a_{m}e^{(A_m-A_{m+1})}\right)
\nonumber\\
&+&
\sum_{p}\omega_{p}b_{p}^{\dagger}b_{p},
%-\sum_{m}\hat n_m \sum_p \alpha_{m,p}^2\omega_{p}.
\label{eq:barHAHbath}
\eea
with the anti-hermitian operator $A_m=\sum_{p} \alpha_{m,p}(b_{p}^{\dagger}-b_{p})$.
The rate constant of electron hopping between neighboring sites can be calculated e.g.,
by treating $v_{m,m'}$ as a small parameter
%a a perturbation which remains small regardless of the strength of the electron-phonon bath coupling
\cite{Silbey}. Recent studies adopted this model for describing coherent
 electronic energy transfer in a protein environment, see for example Refs. \cite{Ex1,Ex2}.

Equation (\ref{eq:HAHbath})
has been often introduced in the literature to model the interaction of electrons or excitons
with a normal-mode environment (phonons, photons), but
a local-anharmonic spin-bath can be similarly implemented.
The $M$-site SF model is given by the Hamiltonian
\bea
H_{SF}^M&=& %\sum_{\nu, k}\epsilon_kc_{\nu,k}^{\dagger}c_{\nu,k} +
\sum_{m}\epsilon_m \hat n_m
+\sum_{m,m'}\left(v_{m,m'}a_{m}^{\dagger}a_{m'} +v_{m',m}^*a^{\dagger}_{m'}a_{m}\right)
\nonumber\\
&+&
\sum_{p}
\frac{\omega_{p}}{2} \sigma_{z}^{p}
+
\sum_m \hat n_m \sum_{p}\alpha_{m,p} \omega_{p}\sigma_x^{p}.
\label{eq:HSFbath}
\eea
The spin bath includes many local modes of spacing $\omega_p$,
 described by the Pauli matrices $\sigma_{x,y,z}^p$, coupled via $\alpha_{m,p}$ to the electronic number operator
on site $m$. This Hamiltonian
can be transformed by extending the procedure of Sec. \ref{ModelII} to receive
%with $A_m=-i\sum_p\lambda_{m,p}\sigma_{y}^{m,p}$
%$\lambda_{m,p}=\frac{1}{2}atan(2\alpha_{m,p})$,
%
\bea
\bar H_{SF}^M&=&
\sum_{m}\epsilon_m \hat n_m
+\sum_{m,m'} \left(v_{m,m'}a_{m}^{\dagger}a_{m'} e^{(A_m-A_{m'})}  + h.c.
%+v_{m+1,m}^* a^{\dagger}_{m+1}a_{m}e^{A_m-A_{m+1}}
\right)
\nonumber\\
&+&
\sum_{p}
\frac{\omega_{p}}{2} \sigma_{z}^{p}
+
\sum_{m} \hat n_m\sum_p \frac{\omega_p}{2}\left(\frac{1-\cos 2\lambda_{m,p}}{\cos2\lambda_{m,p}}\right)
\sigma_{z}^{p}.
\nonumber\\
\label{eq:barHSFbath}
\eea
Here  $\lambda_{m,p}=\frac{1}{2}\arctan(2\alpha_{m,p})$ is
a renormalized coupling parameter
and   $A_m=i\sum_p\lambda_{m,p}\sigma_{y}^{p}$ is
the anti-hermitian operator generating the transformation.
%
%This result was derived once assuming that the electron-spin modes coupling are diluted in the thermodynamics limit
%$\alpha_{m,p}\propto 1/\sqrt{N_{m,p}}$, with $N_{m,p}$ the number of modes in the $m$ bath.
%More genral expressions are included in Ref. \cite{SegalSpin}.
%
It is interesting to extend recent polaron studies of exciton transfer in
biomolecules and examine the dynamics under the local-bath model (\ref{eq:barHSFbath}),
to understand the role of bath harmonicity/anharmonicity (normal modes or local modes)
in sustaining quantum coherent dynamics of electronic degrees of freedom.

%========================

{}


\begin{thebibliography}{5}
\bibitem{commentshift}
Depending on the model, the quantum master equation may be derived by defining a new molecular Hamiltonian
$\bar H'_S=\bar H_S+\rm{tr_B}[\bar H_{SB}]$ and a corresponding interaction term
$\bar H'_{SB}=\bar H_{SB}-\rm{tr_B}[\bar H_{SB}]$. This choice enforces offsetting of the first-order
term in the perturbative expansion, ${\rm tr_B }[\bar H'_{SB}]=0$.
\bibitem{comment1}
Fig. \ref{FigJ1} was generated with the same parameters as Fig. 1(a) Ref. \cite{KochFCB},
%besides the convention for the dot energy $\epsilon_d$.
but here, we set the bare energy as $\epsilon_d=0$,
so as to allow a clear comparison to the SF model.
In contrast, in  Ref. \cite{KochFCB} the renormalized energy was fixed, $\epsilon_d -\alpha^2\omega_0=0$.
Our AH results match Ref. \cite{KochFCB} when we follow their convention.
\end{thebibliography}
\end{document}